\documentstyle[12pt]{article}
\newtheorem{theorem}{Theorem}
\begin{document}
\title{The $2m\leq r$ property of spherically symmetric static spacetimes}
\author{Marc Mars$^\ast$ \\
School of Mathematical Sciences,\\
Queen Mary and Westfield College, \\
Mile End Rd, London, E1 4NS, U.K. \\
and \\
M. Merc\`{e} Mart\'{\i}n-Prats$^\ast$ and 
Jos\'e M. M. 
Senovilla\thanks{Also at Laboratori de F\'{\i}sica Matem\`atica, Societat
Catalana de F\'{\i}sica, IEC, Barcelona.}\\ 
Departament de F\'{\i}sica Fonamental, Universitat de Barcelona\\
 Diagonal 647, 08028 Barcelona, Spain.}
\date{}
\maketitle
\begin{abstract}
We prove that all spherically symmetric static spacetimes which are both
regular at $r=0$ and satisfying the single energy condition
$\rho + p_r + 2p_t \geq 0$ cannot contain any black hole region (equivalently,
they must satisfy $2m/r \leq 1$ everywhere). This result holds even when the
spacetime is allowed to contain a finite number of matching hypersurfaces.
This theorem generalizes a result by Baumgarte and Rendall when the matter
contents of the space-time is a perfect fluid and also complements their
results in the general non-isotropic case.
\end{abstract}


\vspace{1cm}

In a recent paper, Baumgarte and Rendall \cite{tb} have proven that all
spherically symmetric static perfect-fluid solutions of the Einstein field
equations with a regular centre cannot enter into a black hole region
(up to where the pressure vanishes) 
provided that the energy-density $\rho$ is continuous and
non-negative and the central isotropic pressure is finite and positive.
The method for obtaining their result was the analysis
of the Tolman-Oppenheimer-Volkoff equation (we assume $8 \pi G = c = 1$
throughout)
\begin{eqnarray*}
\frac{dp}{dr} = -\frac{ \left (\rho + p \right) \left (2 m(r) +  p r^3
\right )}{2 r^2 \left ( 1- 2 m(r)/r \right )} 
\end{eqnarray*}
where $\rho$ and $p$ are the energy density and pressure of the perfect fluid
and $m(r)$ is the so-called mass function defined by
\begin{equation}
m(r) \equiv \frac{1}{2} \int_{0}^{r}{\rho(v) v^2 dv} . \label{mass}
\end{equation}
Here $r$ stands for the usual Schwarzschild area coordinate (see below).
Their main result was that, in the region where $p$ remains non-negative
(which is the interior of the autogravitating fluid), we have
\begin{equation}
\frac{2m}{r} < 1, \hspace{1cm} \mbox{in the region where} \hspace{3mm}
p(r) \geq 0 ,\label{cond}
\end{equation}
which clearly implies that the spacetime cannot contain
any black hole region (there are no trapped 2-spheres).
This theorem was a generalization of previous results 
\cite{RS} which assumed also the existence of a barotropic equation of state
$p=p(\rho)$ satisfying $dp/d \rho > 0$.

In the same Ref.\cite{tb}, the impossibility of entering into a black hole
region was also shown in the case that the fluid has anisotropic pressures. 
In this case, they proved that condition (\ref{cond}) holds
up to where the radial pressure vanishes (again the interior of the
autogravitating fluid) under the similar assumptions of continuous
and non-negative energy density, positive and finite central
radial pressure and a $C^1$ tangential pressure. The additional condition
they assumed of tangential pressure being coincident with the radial pressure
at $r=0$ is, in fact, a direct consequence of the regularity of the spacetime
at $r=0$.

The aim of this paper is to present a new theorem on spherically
symmetric static spacetimes which is a strict generalization of the theorem
by Baumgarte and Rendall in the perfect-fluid case and that largely complements
their results in the completely general case. In fact, we shall prove by using
very simple arguments that an arbitrary spherically symmetric static spacetime 
satisfying the single condition
\begin{eqnarray}
\rho + p_r + 2 p_t \geq 0, \label{con}
\end{eqnarray}
(where $\rho$, $p_r$ and $p_t$ are defined below) must necessarily fulfil 
\begin{eqnarray*}
\frac{2m}{r} \leq 1 \hspace{1cm}\forall r>0.
\end{eqnarray*}
The only differentiability requirements we impose
are that the matching conditions are fulfilled everywhere (equivalently,
there are no surface layers) and that the spacetime metric is $C^2$ between
any two consecutive matching hypersurfaces. These requirements do not
even imply the continuity of the energy density $\rho$, nor the continuity of
the tangential pressure $p_t$.

In order to prove the theorem, let us choose coordinates $\{t,R,\theta,\phi\}$
in which the metric takes the form
\begin{eqnarray}
ds^2 = - F^2(R) dt^2 + dR^2 + r^2(R) \left( d\theta^2+
\sin^2 \theta d\phi^2 \right), \label{esp} 
\end{eqnarray}
where $r(R)$ is the usual Schwarzschild radius. These coordinates are always
well-defined in at least a neighbourhood of $r=0$ {\it as long as} the
spacetime is regular there. The regularity of the spacetime on a centre of
symmetry $R=0$ (we can fix an additive constant in $R$ in order to set $r(0)=0$,
that is to say, we can choose $R$ such that a regular centre of symmetry is
at $R=0$) is satisfied if and only if the metric functions $r(R)$ and $F(R)$
have the following asymptotic behaviour
\begin{equation}
\displaystyle{r(R \rightarrow 0) \leadsto R - \frac{m_0}{3} R^3 }, \; \; \; 
\displaystyle{F(R \rightarrow 0) \leadsto F_0 + F_1 \, R^2} , 
\; \; \; F_0 >0 \label{reg}
\end{equation}
where $m_0, F_0$ and $F_1$ are constants and $F_0$ is strictly positive.
These conditions will be assumed from now on. This is one of the main
assumptions in our work, as otherwise there is no guarantee that the coordinates
of (\ref{esp}) exist.

The energy-momentum tensor of this spacetime can be calculated
from its Einstein tensor via the Einstein equations. Using the static
velocity vector
\begin{eqnarray*}
\vec{u} = \frac{1}{F} \frac{\partial}{\partial t}
\end{eqnarray*}
to decompose the energy-momentum tensor, we find that the energy-density
$\rho$, radial pressure (or tension) $p_r$ and tangential pressure
(or tension) $p_t$ with respect to $\vec{u}$ are, respectively
\begin{eqnarray*}
\rho = \frac{1}{r^2} \left( 1 - {r'}^2 - 2 r r'' \right), \hspace{5mm}
p_r= \frac{1}{r^2} \left(-1 + {r'}^2 + 2r r' \frac{F'}{F}\right), \hspace{5mm}
p_t =  \frac{F''}{F} + \frac{F'}{F} \frac{r'}{r} + \frac{r''}{r}
\end{eqnarray*}
where the prime denotes derivative with respect to $R$.

As is well-known the mass function $m$ is invariantly defined for
every spherically symmetric spacetime \cite{Zan}, and in our case is given by
the expression
\begin{eqnarray}
2m \equiv r\left(1-r'^2\right), \label{fund}
\end{eqnarray}
from where one can immediately obtain the formula (\ref{mass}) previously
given. In consequence, we have that, {\it as long as} the coordinate system we
have chosen is appropriate, the mass function is bounded to satisfy
\begin{eqnarray*}
\frac{2m}{r} \leq 1 ,
\end{eqnarray*}
the equality holding {\it only} at points with $r'=0$, if any.
Therefore the spacetime does not contain closed trapped surfaces
(which are the indication of the existence of a black hole region) in the
domain where the chosen coordinates are valid. Thus, proving that the spacetime
cannot contain a back hole region under the assumption $\rho+p_r+2p_t \geq 0$
amounts to proving that the coordinates $\{t,R\}$ in which the metric
(\ref{esp}) is written cover the whole manifold
and that this manifold is inextendible.

As is obvious, from expression (\ref{esp}) follows that the coordinates
$\{t,R\}$ will fail describing the spacetime manifold if and only if the metric
function $F(R)$ vanishes somewhere (if the function $F$ diverges for some
$R$, this is either a singularity of the spacetime or we have $r=0$ as can
be easily checked from expression  (\ref{be}) below when considered as a second
order linear  differential equation for $F$). 
It is also a simple matter of checking that if $F(R)$ is everywhere positive
then the manifold is inextendible (all causal curves are inextendible).
As we explained before, the vanishing of $r(R)$ at $R=0$ is just
the indication that we are at the centre of symmetry of the spacetime. The
function $r(R)$ may certainly vanish again at some other value of $R>0$,
but this is either {\it another} regular centre of symmetry or a curvature
singularity of the spacetime.
In both cases, the coordinate system covers the
whole manifold and the inequality $2m\leq r$ will hold everywhere.

From the above we see that,
in order to prove our result, we only have to show that the function $F(R)$
cannot vanish anywhere under the assumptions that the matter content satisfies
the condition (\ref{con}) and that there is {\it a} regular centre of
symmetry $R=0$. The main assumption of our theorem, namely
the condition (\ref{con}), reads explicitly
\begin{eqnarray}     
\frac{F''}{F} + 2 \frac{F'}{F} \frac{r'}{r} =
\frac{1}{2}\left(\rho+p_r+2p_t\right) \geq 0. \label{be} 
\end{eqnarray}     
Consequently, the lefthand side of this relation must be a non-negative
function. This condition can be rewritten as
\begin{eqnarray}
\left ( F' r^2 \right )' =\frac{1}{2}\left(\rho+p_r+2p_t\right) r^2 F.
\label{equa0}
\end{eqnarray}
or, equivalently,
\begin{eqnarray}
F'(R) = \frac{1}{2r^2} \int_{0}^{R}
{\left(\rho+p_r+2p_t\right) r^2 F d\tilde{R} }. \label{equa}
\end{eqnarray} 
It is clear that this equation, together with (\ref{con}) and (\ref{reg}),
implies that $F$ is positive everywhere. This follows immediately from the fact
that $F(0)=F_0>0$, so that there is a non-empty interval $[0,R_1]$
in which $F$ is strictly positive, together with (\ref{equa}) and (\ref{con})
which imply then that $F'(R)$ is non-negative in this interval. Therefore, $F$
is a non-decreasing function in $[0,R_1]$ and thus $F(R_1)\geq F_0 >0$.
Proceeding in this manner (starting now at $R_1$) we see that $F$ is a
non-decreasing function everywhere and in fact we have $F(R)\geq F_0>0$
for all possible $R$.

Now, this proof implicitly assumes that both functions $F(R)$ and $r(R)$
are piecewise $C^2$ (so that the expressions for $\rho$, $p_r$ and $p_t$
make sense) and that the function $F$ is $C^1$ everywhere (otherwise we
are not allowed to integrate (\ref{equa0}) to give simply (\ref{equa})).
Our aim now is to show that this differentiability requirements
are fulfilled if the spacetime under consideration 
contains a finite number of matching hypersurfaces. Let us here
emphasize that, by allowing for the existence of those matching hypersurfaces,
we are including in our treatment spacetimes representing not only the 
interior of a star but also its exterior, which can be either empty
or filled with some exterior field such as, for example, an electromagnetic
field. Notice that in all the previous theorems (e.g. in \cite{tb})
only the region up to the limit surface of the star was considered.

By assuming the absence of surface layers, the junction conditions reduce to
the equality of both the first and the second fundamental forms calculated from 
the two sides of the matching hypersurface (see e.g. \cite{MS}\cite{BV}). Let us
remark here that, in general, these conditions do not imply the continuity of
the first derivatives of the metric coefficients in a prefixed coordinate
system (see, for instance, \cite{MS}\cite{BV}). Nevertheless, a straightforward
calculation shows that, in our case, the junction conditions on the matching
hypersurfaces $R= \bar R = \mbox{const.}$ reduce simply to 
\begin{eqnarray*}
\left. F \right|_{R \rightarrow \bar{R}^{+}} =
\left.F \right|_{R \rightarrow \bar{R}^{-}}, \hspace{1cm}
\left. r \right|_{R \rightarrow \bar{R}^{+}} =
\left. r \right|_{R \rightarrow \bar{R}^{-}}, \\
\left. F' \right|_{R \rightarrow \bar{R}^{+}} =
\left. F' \right|_{R \rightarrow \bar{R}^{-}}, \hspace{1cm}
\left. r' \right|_{R \rightarrow \bar{R}^{+}} =
\left. r' \right|_{R \rightarrow \bar{R}^{-}},
\end{eqnarray*}
so that both functions $F(R)$ and $r(R)$ have continuous derivatives
everywhere and the result is established. We may notice by the way that this
result means that the coordinate system $\{t,R,\theta,\phi\}$ of (\ref{esp})
is {\it admissible} in the sense of Lichnerowicz \cite{BV}\cite{L}.
We have thus established the following theorem.
\begin{theorem}		
In an arbitrary static and spherically symmetric spacetime 
which is piecewise $C^2$ and satisfies the
matching conditions on a finite number of hypersurfaces, if the
spacetime is regular at a centre of symmetry and
satisfies $\rho + p_r + 2 p_t \geq 0$, then $2m\leq r$ everywhere.
\end{theorem}
Let us remark that this result has been proved irrespective
of the matter content of the spacetime
(provided that the condition $\rho+p_r + 2 p_t \geq 0$ holds everywhere).
Thus, the energy-momentum tensor can certainly represent a real static fluid,
but also more complicated situations in which matter and/or additional fields
(typically electromagnetic fields) are present. The case with empty regions
is also covered by our treatment.

On the other hand, let us notice that the strict inequality $2m/r < 1$ holds
everywhere except at points where $r' = 0$, as we already remarked after
Eq.(\ref{fund}). However, at those points, the expression for $p_r$ becomes
\begin{eqnarray*}
\left. p_r \right|_{r'=0} = - \frac{1}{r^2} < 0.  
\end{eqnarray*}
Thus, the equality $2m = r$ is forbidden as long as $p_r$ remains positive.
In other words, the strict inequality $2m<r$ of Ref.\cite{tb} is fully recovered 
because in \cite{tb} this was shown only up to where $p_r$ vanishes.
It may be interesting to remark that the possible hypersurfaces with $r'=0$ (or 
equivalently $2m=r$) are completely ordinary {\it timelike} hypersurfaces
if $F$ does not vanish there (and $F$ does not vanish if (\ref{con}) and
(\ref{reg}) hold, as we have seen). Therefore, the possible hypersurfaces
$2m=r$ which appear in our treatment do not have anything to do with
horizons or other null hypersurfaces, and the Killing vector
$\partial /\partial t$ is timelike everywhere if condition (\ref{con}) holds.

To end this letter, let us briefly comment on the
main condition (\ref{con}) we have assumed in the theorem. This condition
is implied by (but is much less restrictive than) the strong energy
condition (see, e.g., \cite{he}). For the spherically symmetric
static case we are dealing with, the strong energy condition is equivalent to
\begin{eqnarray*}
\rho + p_r \geq 0, \hspace{2cm} \rho + p_t \geq 0 ,\hspace{2cm}
\rho + p_r + 2 p_t \geq 0.
\end{eqnarray*}
Thus, our condition (\ref{con}) is {\it just} one of the consequences of the
strong energy condition.

The present work has been partially supported by the 
Spanish Ministerio de Educaci\'{o}n y Ciencia under project
No.\ PB93-1050. M. Mars wishes to thank the Direcci\'o Ge\-ne\-ral
d'Universitats, Generalitat de Catalunya, for finantial support.

\end{document}